\documentclass{article}
\usepackage{graphicx}
\usepackage{amssymb}
\usepackage{amsmath}
\usepackage{xcolor}
\usepackage[colorlinks=true ,urlcolor=blue ,anchorcolor=blue ,citecolor=blue ,filecolor=blue ,linkcolor=blue ,menucolor=blue ,pagecolor=blue ,linktocpage=true ,pdfproducer=medialab ,pdfa=true]{hyperref}
\usepackage{jcappub}
\usepackage{mathrsfs}
\usepackage[utf8]{inputenc}
\usepackage{color,bm,comment,braket}
\usepackage[T1]{fontenc}
\usepackage[normalem]{ulem}

\title{Revisiting Affleck-Dine Leptogenesis with light sleptons}
\author[a]{Kazuki Enomoto,}
\author[b,c]{Koichi Hamaguchi,}
\author[d]{Kohei Kamada,} 
\author[b]{Juntaro Wada}

\affiliation[a]{Department of Physics, KAIST, Daejeon 34141, Korea}
\affiliation[b]{Department of Physics, University of Tokyo, Bunkyo-ku, Tokyo 113--0033, Japan}
\affiliation[c]{Kavli IPMU (WPI), University of Tokyo, Kashiwa, Chiba 277--8583, Japan}
\affiliation[d]{Research Center for the Early Universe (RESCEU), Graduate School of Science, The University of Tokyo, Hongo 7-3-1 Bunkyo-ku, Tokyo 113-0033, Japan}
\emailAdd{k\_enomoto@kaist.ac.kr, 
hama@hep-th.phys.s.u-tokyo.ac.jp, 
kohei.kamada@resceu.s.u-tokyo.ac.jp, 
wada@hep-th.phys.s.u-tokyo.ac.jp}
\date{\today}

\abstract{We revisit the Affleck-Dine leptogenesis via the $L H_u$ flat direction with a light slepton field.
Although the light slepton field is favored in low-energy SUSY phenomenologies, such as the muon $g-2$ anomaly and bino-slepton coannihilation, it may cause a problem in the Affleck-Dine leptogenesis: 
it may create an unwanted charge-breaking vacuum in the Affleck-Dine field potential
so that the Affleck-Dine field is trapped during the course of leptogenesis. We investigate the conditions under which such an unwanted vacuum exists and 
clarify that both thermal and quantum corrections are important for the (temporal) disappearance of the charge-breaking minimum. We also confirm that if the charge-breaking vacuum disappears due to the thermal or quantum correction, the correct baryon asymmetry can be produced while avoiding the cosmological gravitino problem.
}

\begin{document}

\begin{flushright}
RESCEU-8/23
\end{flushright}

\maketitle

\section{Introduction}
Leptogenesis~\cite{Fukugita:1986hr} is one of the most attractive scenarios to explain the origin of the baryon asymmetry of the Universe. In this scenario, the matter-antimatter asymmetry is first generated in the form of lepton asymmetry via lepton-number violating interactions, which are motivated by the non-zero neutrino masses~\cite{ParticleDataGroup:2022pth} via the seesaw mechanism~\cite{Minkowski:1977sc,Yanagida:1979as,Yanagida:1980xy,Gell-Mann:1979vob}. The generated lepton asymmetry is then converted into the baryon asymmetry via the sphaleron effect~\cite{Kuzmin:1985mm}. Among the various realizations of the scenario, the leptogenesis via the $L H_u$ flat direction~\cite{Murayama:1993em}, based on the Affleck-Dine (AD) mechanism~\cite{Affleck:1984fy}, is an interesting possibility in the framework of supersymmetry (SUSY). 
In this scenario, the slepton field plays the central role as a constituent of the AD field and as a carrier of the lepton number.  In this work, we revisit this scenario, AD leptogenesis, with the light slepton, favored for the low-energy SUSY phenomenology.

SUSY searches at the LHC have limited the masses of the squarks to be $O(1)~\mathrm{TeV}$ or heavier, while the sleptons can still be $O(100)~\mathrm{GeV}$~\cite{ATLAS:2023lon,CMS_SUSY}.
Such a light slepton is favored in the low-energy SUSY phenomenologies. For example, in the context of the bino dark matter, the light slepton can help reduce its abundance by the coannihilation to realize the correct relic dark matter density~\cite{Ellis:1998kh,Ellis:1999mm,Baer:2002fv,Nihei:2002sc,Edsjo:2003us}. In the context of the muon $g-2$ anomaly,\footnote{The experimentally measured value of the muon $g-2$~\cite{Muong-2:2006rrc,Muong-2:2021ojo} deviates from the prediction of the Standard Model of particle physics (SM)~\cite{Aoyama:2020ynm}, where the hadronic vacuum polarization (HVP) contribution is obtained in data-driven methods. Recent results from lattice QCD calculations for the HVP contribution~\cite{Borsanyi:2020mff,Blum:2023qou, Alexandrou:2022tyn} indicate a significantly smaller deviation, and a recent new measurements of $\sigma(e^+e^-\to \mathrm{hadron})$~\cite{CMD-3:2023alj} also hints at a smaller HVP contribution than the one using previous experimental inputs in the data-driven methods. The situation is still controversial and the reasons for the deviations are not clear.}
the light slepton plays a crucial role in explaining this anomaly because the sleptons interact with leptons at the tree level~\cite{Lopez:1993vi, Chattopadhyay:1995ae, Moroi:1995yh,Endo:2021zal}.
Since the slepton is part of the scalar field along the $LH_u$ flat direction, or the AD field, the light slepton can lead to a light AD field.

Such a light AD field, however, may cause a serious problem in AD leptogenesis.
This is because if the AD field is lighter than the gravitino, a charge-breaking vacuum (CBV) appears along the flat direction~\cite{Kawasaki:2000ye}.
Even if the electroweak vacuum is classically stable and has a lifetime longer than the age of our Universe, the AD field may eventually fall into the charge-breaking minimum because it has a large field value during the AD mechanism.
One may wonder if the CBV problem is unavoidable so that the AD leptogenesis via the $L H_u$ flat direction
does not work for TeV-scale or higher-scale gravitinos, which would severely constrain the SUSY mass spectrum. Indeed, the CBV problem in the AD baryogenesis
has been discussed in Ref.~\cite{Kawasaki:2006yb} in the context of the anomaly-mediated SUSY-breaking models~\cite{Randall:1998uk,Giudice:1998xp}.

In this work, we revisit this problem with the light slepton field and discuss the condition under which this unwanted vacuum exists. We show that both thermal and quantum corrections are important for the (temporal) disappearance of the charge-breaking minimum: the thermal logarithmic potential and the running effect of the AD field mass. We also investigate the condition under which the AD field is not trapped by the unwanted minimum and confirm that the correct baryon asymmetry can be produced while also avoiding the cosmological gravitino problem~\cite{Khlopov:1984pf,Ellis:1984eq,Kawasaki:2008qe,Kawasaki:2017bqm}.

In addition, we discuss the connection between the AD leptogenesis and low-energy SUSY phenomenology.
We find that the running effect is important to hide the CBV when the slepton is sufficiently lighter than the squark.
In such a case, the produced asymmetry becomes almost independent of the light slepton mass in the low-energy SUSY model, due to the running effect. 

The paper is organized as follows. In Sec.~\ref{sec:ADLG}, after a brief review of the AD leptogenesis, we discuss the appearance of the unwanted CBV for light sleptons. In Sec.~\ref{sec:solution}, we discuss two solutions to avoid the CBV problem, the running and the thermal effects. We show our results in Sec.~\ref{sec:result}, and Sec.~\ref{sec:summary} is devoted to a summary.

\section{Affleck-Dine leptogenesis with light slepton}
\label{sec:ADLG}
In this section, we first briefly review the Affleck-Dine leptogenesis in Sec.~\ref{subsec:ADLG}. We then discuss in Sec.~\ref{subsec:CBV} that the light slepton mass can cause a problematic CBV along the AD potential.

\subsection{Brief review of Affleck-Dine leptogenesis}
\label{subsec:ADLG}

The existence of flat directions along which the scalar potential vanishes
is one of the cosmologically important characteristics of the supersymmetric theories~\cite{Affleck:1984fy}, 
since the scalar field can easily acquire a large expectation value in the early Universe. 
In the supersymmetric SM, the flat directions are classified by gauge invariant polynomials~\cite{Gherghetta:1995dv}. 
Among them, the $LH_u$ flat direction in the minimal supersymmetric SM (MSSM) is parameterized as~\cite{Murayama:1993em}
\begin{align}
    \widetilde{L} &= 
    \frac{1}{\sqrt{2}}\begin{pmatrix}\phi\\0 \end{pmatrix},
    \quad
    H_u = 
    \frac{1}{\sqrt{2}}\begin{pmatrix}0\\\phi \end{pmatrix},
\end{align}
where $\widetilde{L}$ and $H_u$ are the left-handed slepton and the up-type Higgs field, respectively. 
We call the $\phi$ field parameterizing the flat direction as the AD field.
This $LH_u$ flat direction is particularly interesting because it is directly related to the neutrino masses through the lepton number-violating operator~\cite{Murayama:1993em}
\begin{align}
W &= \frac{1}{2M}(LH_u)^2 = \frac{1}{8M}\phi^4\,,
\label{eq:LHLH}
\end{align}
where $L$, $H_u$, and $\phi$ denote the corresponding superfields, and $M$ is a parameter with mass dimension one.  This operator leads to a neutrino mass
\begin{align}
m_\nu &= \frac{v_u^2}{M} \simeq 3\times 10^{-9} \mathrm{eV} \times \left(\frac{M}{10^{22}~\mathrm {GeV}}\right)^{-1} \sin^2 \beta, 
\label{eq:M_and_neutrino_mass}
\end{align}
where $v_u = \langle H_u \rangle \simeq 174 \mathrm{GeV} \times  \sin \beta$ with $\tan \beta \equiv  \langle H_u \rangle/\langle H_d \rangle$ ($H_d$ is the down-type Higgs field). 
Hereafter, we assume that the neutrino masses are hierarchical and focus on a single flat direction corresponding to the lightest active neutrino.
The value of $M$ can be larger than the reduced Planck mass because it is an effective parameter
that can involve, {\it e.g.}, a small Yukawa coupling, as in the type-I see-saw mechanism.

The scalar potential for the AD field $\phi$ is given by
\begin{align}
V_0(\phi) &= m_{\phi}^2 |\phi|^2 + \left(a_m\frac{m_{3/2}\phi^4}{8M}+h.c.\right)+\frac{|\phi|^6}{4M^2}.
\label{eq:V0}
\end{align}
The mass $m_\phi$ is given by
\begin{align}
    m_{\phi}^2 & = \frac{1}{2}\left( m_{\tilde{L}}^2 + m_{H_u}^2 +\mu^2\right),
    \label{eq:mphi}
\end{align}
where $m_{\tilde{L}}^2$, $m_{H_u}^2$ are the slepton and the up-type Higgs mass squared, respectively, and $\mu$ is the Higgsino mass parameter. 
The second term (A-term) in Eq.~\eqref{eq:V0} comes from the SUSY-breaking effect in supergravity and is proportional to the gravitino mass, $m_{3/2}$. The coefficient $a_m$ is expected to be $O(1)$, and we take it to be real and positive by field redefinitions. This term comes from the operator in Eq.~\eqref{eq:LHLH} and violates the lepton number.
The third term in Eq.~\eqref{eq:V0} is the F-term potential induced by the superpotential~\eqref{eq:LHLH}.

Next, let us consider the cosmological dynamics of the AD field.
In the early Universe, the potential is modified as follows.
\begin{align}
V &= V_0
+ V_\mathrm{HM}
+ V_{\mathrm{thermal}},
\label{eq:Vtotal}
\end{align}
where $V_0$ is given by Eq.~\eqref{eq:V0}.
The second term is the Hubble-induced mass term induced by additional SUSY-breaking effects from the inflaton sector~\cite{Dine:1995uk,Dine:1995kz},\footnote{We have neglected the Hubble-induced A-term, since it is hard to generate it in most of the supergravity F-term inflation without having severe fine-tunings or cosmological difficulties~\cite{Kasuya:2008xp,Kamada:2008sv}.}
\begin{align}
V_\mathrm{HM} &= -c H^2 |\phi|^2, 
\end{align}
where $H$ is the Hubble parameter during and after the inflation. The coefficient $c$ is a model-dependent $O(1)$ parameter, and we assume $c>0$.
The third term in Eq.~\eqref{eq:Vtotal} represents the thermal effects and is given by~\cite{Allahverdi:2000zd,Anisimov:2000wx}
\begin{align}
V_{\mathrm{thermal}} &\simeq 
\sum_{f_k |\phi| < T} c_k f_k^2 T^2 |\phi|^2
+
a_g \alpha_{3}^2 T^4 \log{\left(\frac{|\phi|^2}{T^2}\right)}.
\end{align}
Here, $T$ is the temperature,
$f_k$ denote gauge and Yukawa couplings relevant for the $LH_u$ flat direction with $c_k$ being constants of $O(1)$~\cite{Asaka:2000nb},
$a_g=1.125$~\cite{Hamaguchi:2002vc}, and $\alpha_{3}=g_3^2/(4\pi)$ with $g_3=g_3(T)$ being the SU(3) running coupling.
Note that thermal plasma appears in the Universe
even before the reheating, as a subdominant component of the Universe, with a temperature $T\simeq (T_R^2 M_{p} H)^{1/4}$~\cite{Kolb:1990vq}, where $T_R$ and $M_p$ are the reheating temperature and the reduced Planck scale, respectively.
 
When the Hubble parameter is sufficiently large, $H \gg m_\phi$, the AD field rolls along a time-dependent potential minimum, $|\phi| \sim (H M)^{1/2}$, 
which is determined by the balance between the Hubble-induced mass term and the $|\phi|^6$ term.
Eventually, one of the 
other terms in Eq.~\eqref{eq:Vtotal}
becomes comparable to the Hubble-induced mass, and
the AD field begins to oscillate coherently. Such an epoch is determined by~\cite{Fujii:2001zr}
\begin{align}
H^2 \simeq m_\phi^2 +\sum_{f_k |\phi| < T} c_k f_k^2 T^2 
+ a_g \alpha_{3}^2 \frac{T^4}{|\phi|^2},
\end{align}
which leads to 
\begin{align}
\label{eq: Hoscdefinition}
H_{\mathrm{osc}} 
&\simeq 
\mathrm{max}
\left[
m_\phi,\,
H_k,\,
\alpha_3 T_R \left(\frac{a_g M_p}{M}\right)^{1/2}
\right],
\\
H_k &= \mathrm{min}\left[\frac{M_p T_R^2}{f_k^4 M^2}, (c_k^2 f_k^4 M_p T_R^2)^{1/3}\right].
\end{align}
At the same time, the lepton-number-violating 
A-term gives a kick to the phase direction of the AD field
with an efficiency of $O(m_{3/2}/H_\mathrm{osc})$
so that the angular momentum in the field space (or equivalently, lepton number) is generated 
in the Universe.
Soon after the onset of the oscillation, 
the A-term becomes irrelevant, and the comoving
lepton asymmetry is fixed. 
This asymmetry is eventually
converted to the fermion sector through the decay/melting of the scalar condensate and to the
baryon asymmetry through the sphaleron process~\cite{Kuzmin:1985mm}. 
Then, the net baryon-to-entropy ratio 
is fixed at reheating and
can be estimated as~\cite{Fujii:2001zr}
\begin{align}
\label{eq:baryonasymmetry}
\frac{n_B}{s} &=\frac{2}{69} \delta_{\mathrm{CP}}\frac{a_m m_{3/2}T_{R}}{H_{\mathrm{osc}}M_{p}}\left(\frac{M}{M_{p}}\right),
\end{align}
where $\delta_{\mathrm{CP}}\sim O(1)$ represents the phase of CP violation determined by the initial phase of the AD field during inflation.

\subsection{Light slepton mass and charge-breaking vacuum}
\label{subsec:CBV}

Let us now discuss the problem caused by the light slepton. In the case of $LH_u$ flat direction, the AD field mass consists of the slepton mass and the sum of the up-type Higgs mass and the $\mu$-term, as we have explained in Eq.~\eqref{eq:mphi}. The part other than the slepton mass is given by~\cite{Martin:1997ns}
\begin{align}
m_{H_u}^2 + \mu^2 
&=
-\frac{m_Z^2}{2} 
+\frac{1}{2}\left(
\frac{\left|m_{H_d}^2-m_{H_u}^2\right|}{\sqrt{1-\sin^2(2\beta)}}-(m_{H_d}^2-m_{H_u}^2)
\right)
\\&=
-\frac{m_Z^2}{2} 
+\frac{m_{H_d}^2-m_{H_u}^2}{\tan^2\beta-1},
\label{higgsandmuterm}
\end{align}
where $m_{H_d}^2$ is the down-type Higgs mass squared, and we have assumed $m_{H_u}^2<m_{H_d}^2$ and $\tan\beta>1$ in the second line, which is typically the case in the MSSM~\cite{Martin:1997ns}. Therefore, if the second term is $O(100)~\mathrm{GeV}$ or smaller,\footnote{
\label{muong_2tanbeta}
In particular, the models that can explain the muon $g-2$ anomaly typically have
large $\tan\beta$~\cite{Lopez:1993vi, Chattopadhyay:1995ae, Moroi:1995yh,Endo:2021zal}.
The second term in Eq.~\eqref{higgsandmuterm} is suppressed in such models. 
} a slepton mass of the order of the electroweak scale leads to sub-TeV AD mass, $m_\phi \sim O(100)~\mathrm{GeV}$. We will focus on this situation in the following.

Then, such a light AD field mass is problematic because it can cause a charge-breaking minimum along the flat direction. 
 Indeed, if the condition
\begin{align}
\label{eq:conditionofexisteceCBV}
m_\phi < \frac{1}{\sqrt{12}}a_m m_{3/2},
\end{align}
is satisfied, the potential~\eqref{eq:V0} has a local charge-breaking minimum, which is given by
\begin{align}
\label{eq:unwantedvaccume}
V(\phi_{\mathrm{min}})
&=
-\frac{f_\phi^2 (4f_\phi-3)}{27}  
a_m^3 m_{3/2}^3 M,
\\
|\phi|_{\rm min} &= \left(\frac{2f_\phi}{3} a_m m_{3/2} M\right)^{1/2},
\end{align}
where
\begin{align}
\label{eq:fphidefinition}
f_\phi = \frac{1}{2}\left(1 + \sqrt{1-\frac{12m_\phi^2}{a_m^2 m_{3/2}^2}}\right).
\end{align}
For $m_{\phi} < (a_m/4) ~m_{3/2}$, it becomes the global minimum, i.e. $V(\phi_{\mathrm{min}})<0$.
In the limit of heavy gravitino, $m_{3/2}\gg m_\phi$, it is reduced to~\cite{Kawasaki:2000ye,Kawasaki:2006yb}
\begin{align}
V(\phi_{\mathrm{min}})&\simeq -\frac{1}{27}a_m^3 m_{3/2}^3 M,\\
|\phi_{\mathrm{min}}|&\simeq \left(\frac{2}{3}a_m m_{3/2}M\right)^{\frac{1}{2}}.
\end{align}
The existence of this unwanted vacuum is problematic because the AD field initially has a large field value. Even if the electroweak vacuum is classically stable and its lifetime is longer than the age of our Universe, the AD field may 
fall into the charge-breaking minimum during the course of the AD mechanism, 
which is inconsistent with the present Universe.

\section{Solutions to the charge-breaking vacuum problem}
\label{sec:solution}
In this section, we discuss the two ways of avoiding the charge-breaking vacuum. The first one is the running effect of the AD field mass, and the second is the thermal effect, which mainly comes from the thermal logarithmic potential. Since these effects can change the shape of the effective potential, they play an important role in hiding the charge-breaking vacuum.

\subsection{Running effect of the AD field}
\label{subsec:running}
In the previous section, we showed that Eq.~\eqref{eq:mphi} indicates $m_\phi\simeq O(100)$ GeV at the {\it electroweak} scale, and that such a light AD field is problematic because it may create the CBV in the AD leptogenesis scenario. However, this expectation overlooked the running effect of the AD field mass. 

In fact, we should note that the existence of the charge-breaking minimum is determined by the shape of the effective potential. This means that the mass of the AD field should be interpreted as the effective mass in our context. To clarify this point, hereafter, we replace it as 
\begin{align}
m_{\phi}^2 \to m_{\phi,\mathrm{eff}}^2:= m_{\phi}^2+\delta m_{\phi}^2,
\end{align}
where $m_{\phi}$ is the AD field mass at the tree level, and $\delta m_{\phi}$ denotes the quantum correction. For simplicity, we only consider the contribution from the one-loop correction, which is proportional to $\ln{\phi}$. 
Then, the quantum correction term $\delta m_{\phi}^2\sim \ln \phi$ is important to determine the existence of the charge-breaking minimum since the AD field takes a large field value when the lepton asymmetry is generated.

As we have explained, the AD field mass includes the slepton mass, up-type Higgs mass, and $\mu$-term.
\begin{align}
    m_{\phi,\mathrm{eff}}^2 & = \frac{1}{2}\left( m_{\tilde{L}}^2 + m_{H_u}^2 +\mu^2\right).
\end{align}
In particular, the running effect is mainly coming from the up-type Higgs mass. Hence, let us investigate the renormalization group equation (RGE) for this.  The up-type Higgs soft mass, $m_{H_u}^2$, receives\footnote{We omit the contribution from the SUSY-breaking sector, which could be significant in the case of gauge mediation~\cite{deGouvea:1997afu}. Note that 
in our case with $m_{3/2} \gtrsim m_{\phi}$, gauge mediation is unlikely the origin of the SUSY breaking.},  in particular, a large contribution from the stop masses~\cite{Martin:1997ns}:
\begin{align}
    \frac{d}{d\ln Q}m_{H_u}^2 
    \sim
\frac{1}{16\pi^2}
    6|y_t|^2 (m_{Q_3}^2 + m_{u_3}^2),
\end{align}
where $y_t$ is the top Yukawa coupling, $m_{Q_3}$ and $m_{u_3}$ are the soft masses of the left-handed and up-type right-handed squarks, respectively. We have kept only the dominant stop contribution and assumed that $A$-term contribution is
small, for simplicity. 
Thus, the running up-type Higgs mass can be approximated by
\begin{align}
\delta  m_{H_u}^2(Q)
&\sim
\frac{12|y_t|^2}{16\pi^2} m_{\tilde{t}}^2
     \ln\left(\frac{Q}{m_{\tilde{t}}}\right),
\end{align}
where $Q$ denotes the running scale, which corresponds to the AD field value.
Here, we assume $m_{Q_3}\sim m_{u_3}\sim m_{\tilde{t}}$, for simplicity.
Then, we obtain the field-value dependent quantum correction to the AD field mass as
\begin{align}
\delta m_{\phi}^2 (\phi)
&\simeq \frac{12|y_t|^2}{32\pi^2} m_{\tilde{t}}^2 \ln\left(\frac{\phi}{m_{\tilde{t}}}\right)
\\
&\sim
(3~\mathrm{TeV})^2
\left(\frac{m_{\tilde{t}}}{3~\mathrm{TeV}}\right)^2
\left(1+\frac{1}{22}\ln\left({3~\mathrm{TeV}\over m_{\tilde{t}}}\frac{\phi}{10^{13}\mathrm{GeV}}\right)\right).\label{deltamphilog}
\end{align}
Here  we take $y_t\sim 1$ and $m_{\tilde{t}}\sim 3$ TeV as a typical value for the stop mass, because $m_{\tilde{t}} \gtrsim O(\mathrm{TeV})$ is motivated by the mass of the SM-like Higgs boson as well as by the LHC constraints.
The AD field value $\phi \sim 10^{13}$ GeV is the typical scale for the AD field during AD leptogenesis (cf.~Fig.~\ref{fig:phioscandcutoff}). Therefore, the AD field mass at the oscillation epoch is 
dominated by the quantum correction to the up-type Higgs mass and is given by
\begin{align}
m_{\phi,\mathrm{eff}} \simeq 3~\mathrm{TeV},
\end{align}
 even if $m_{\phi,\mathrm{eff}}$ is below the $O(1)~\mathrm{TeV}$ at the electroweak scale. This means that during the AD leptogenesis, the AD field mass is almost independent of the light slepton mass.

The existence of the CBV depends on the balance between the mass term and the gravitino A-term in the potential~\eqref{eq:V0}. The running effect of the coefficient of the gravitino A-term, $a_{m} m_{3/2}$, is sufficiently small~\cite{Davidson:2006tg}. 
Therefore, the condition~\eqref{eq:conditionofexisteceCBV} is rewritten as 
\begin{align}
m_{\phi,\mathrm{eff}} < \frac{1}{\sqrt{12}}a_m m_{3/2},
\label{eq:new_CBV_condition}
\end{align}
after we include the running effect.\footnote{
Here, for simplicity, we have neglected the contribution from the derivative of the $m_{\phi,\mathrm{eff}}^2 \sim \ln \phi$ when evaluating the global minimum of the potential, which is sufficiently small as far as $\ln(\phi/m_{\tilde{t}})\gg 1$.
} 
This means that the light slepton will not cause the CBV problem if the gravitino mass is smaller than the stop mass. This running effect is important when the slepton is sufficiently lighter than the stop.

\subsection{Thermal effect}
Next, let us discuss the thermal effect. 
Suppose that the condition~\eqref{eq:new_CBV_condition} is satisfied.
Then, the charge-breaking vacuum exists at a low  temperature, which may spoil the AD leptogenesis scenario. However, if the reheating temperature is high enough, the thermal 
potential can hide the unwanted valley of the potential (\ref{eq:unwantedvaccume}). This point was discussed in Ref.~\cite{Kawasaki:2006yb} in the context of the anomaly-mediation model. 

Let us confirm this argument from the viewpoint of the dynamics of the AD field. Before the oscillation, the AD field follows the potential minimum $|\phi| \simeq \sqrt{HM}$. In the successful AD leptogenesis scenario, the AD field begins to oscillate when $H\simeq H_{\mathrm{ osc}}$ and the motion along the phase direction is also induced at about this time. However, if the curvature along the phase direction becomes larger than the Hubble parameter before the onset of the oscillation, the AD field can quickly fall into the unwanted CBV. Therefore, 
the condition to avoid the CBV problem is described by
\begin{align}
\left|V_{\mathrm{phase}}''\right|_{H=H_{\mathrm{osc}}} \lesssim H^2_{\mathrm{osc}},
\end{align}
where $V_{\mathrm{phase}}''=\partial^2 V/\partial \varphi^2$ with $\varphi$ defined by $\phi = |\phi|\exp(i\varphi/\sqrt{2}|\phi|)$.
The above condition is roughly evaluated as the lower bound on $H_{\mathrm{osc}}$
\begin{align}
a_m m_{3/2} \lesssim H_{\mathrm{osc}},
\label{eq:condition_thermal}
\end{align}
where $H_{\mathrm{osc}}$ is given by Eq.~\eqref{eq: Hoscdefinition}. In particular, when $H_{\mathrm{osc}}$ is determined by the thermal logarithmic potential, this inequality is reduced to
\begin{align}
\frac{a_m m_{3/2}}{\alpha_3}\left(\frac{M}{a_g M_{p}}\right)^{1/2}  \lesssim T_R, \label{TRconstThermalLog}
\end{align}
which is the same expression as in Ref.~\cite{Kawasaki:2006yb}\footnote{
In Ref.~\cite{Kawasaki:2006yb}, the following condition
\begin{align}
a_g \alpha_{3}^2 T_{\mathrm{osc}}^4 > |V(\phi_{\mathrm{min}})|,
\end{align}
is required, which means that the unwanted vacuum must vanish when the AD field starts to oscillate due to the thermal logarithmic potential.}. This condition implies that a sufficiently high reheating temperature is required to avoid the charge-breaking vacuum.

\subsection{Condition to avoid the charge-breaking vacuum}
Before proceeding, we summarize the conditions under which the AD field does not fall into a charge-breaking vacuum. A sufficient condition is
\begin{align}
\frac{1}{\sqrt{12}}a_m m_{3/2} < m_{\phi,\mathrm{eff}},
\label{eq:sufficient}
\end{align}
which means that the charge-breaking local minimum does not exist. As we have seen, even if the AD field mass is small at the electroweak scale, the effective mass may exceed the gravitino mass during AD leptogenesis. Therefore, the above condition should be evaluated at the scale corresponding to the AD field value. Conversely, if this condition is not satisfied, a sufficient thermal effect is required at the oscillation epoch; otherwise, the AD field falls into the CBV.

Here, we note that within the following range
\begin{align}
\frac{1}{4}a_m m_{3/2} < m_{\phi,\mathrm{eff}} < \frac{1}{\sqrt{12}}a_m m_{3/2},
\end{align}
the potential does not have a charge-breaking global minimum but has a local minimum at zero temperature. In this range, the AD field might be trapped by the minimum because the AD field has a large field value during the AD leptogenesis. Even if the AD field finally comes back to the electroweak vacuum through the tunneling effect, the produced baryon asymmetry will be highly suppressed once the AD field is trapped. Therefore, a sufficient thermal effect should be required in this case. If the reheating temperature is not sufficiently high, the situation becomes subtle. At least, we cannot rely on the usual expression of the produced baryon asymmetry in Eq.~\eqref{eq:baryonasymmetry}.

As we have seen in the previous subsection, the condition at the oscillation epoch is given by
\begin{align}
a_m m_{3/2} \lesssim H_{\mathrm{osc}}.
\end{align}
Actually, this inequality is not a necessary condition to avoid the unwanted vacuum. This is because if $H_{\mathrm{osc}}$ is determined by the AD field mass $m_{\phi,\mathrm{eff}}$, then the above condition is reduced to
\begin{align}
a_m m_{3/2} \lesssim m_{\phi,\mathrm{eff}},
\end{align}
and it is stronger than the sufficient condition \eqref{eq:sufficient}.
Therefore, there is a subtle region where the AD field mass is within the following range,
\begin{align}
\frac{1}{\sqrt{12}}a_m m_{3/2} < m_{\phi,\mathrm{eff}} < a_m m_{3/2}.
\end{align}
In this range, 
the AD field is trapped by the unwanted local minimum induced by the negative Hubble-induced mass at the oscillation epoch, if the thermal correction is not large enough to hide this minimum. However, it will eventually come back to the electroweak vacuum as the unwanted local minimum disappears as the Hubble parameter decreases. 
In this case, the produced baryon asymmetry may be reduced due to the temporal trapping, but we expect this reduction to be insignificant in this case since this vacuum will disappear soon after the AD field begins to oscillate.

In conclusion, AD leptogenesis scenarios with light AD fields can be classified into four cases, as shown in Table.~\ref{tab:ADLGscnario}. Case I corresponds to the usual AD leptogenesis scenario because the potential does not have a CBV. In Case II, although the CBV exists in the potential at zero temperature, the AD field is not trapped by this vacuum, due to the thermal correction. In Case III, the potential does not have an unwanted global minimum at zero temperature, but during the leptogenesis, the AD field might be trapped by the unwanted local minimum. 
In the range of $a_m m_{3/2} < \sqrt{12} ~m_{\phi,\mathrm{eff}}$, this local minimum disappears as the Hubble parameter decreases, whereas it remains for $\sqrt{12} ~m_{\phi,\mathrm{eff}} < a_m m_{3/2}  < 4 m_{\phi,\mathrm{eff}}$.
In both cases, the baryon asymmetry should be reduced due to the trapping, and the reduction is expected to be larger as the gravitino becomes heavier. Finally, in Case IV, since the AD field is trapped and falls into the CBV, the AD leptogenesis mechanism does not work.

\begin{table}[t]
    \renewcommand{\arraystretch}{2}
    \centering
    \begin{tabular}{|c|c|c|c|}
    \hline
         & $\displaystyle{\frac{a_m m_{3/2}}{m_{\phi,\mathrm{eff}}}} < \sqrt{12} $ & $ \sqrt{12} < \displaystyle{\frac{a_m m_{3/2}}{m_{\phi,\mathrm{eff}}}} < 4   $ & $4 < \displaystyle{\frac{a_m m_{3/2}}{m_{\phi,\mathrm{eff}}}}$
         \\ \hline
         $H_{\mathrm{osc}} > a_m m_{3/2}$ & 
         I: CBV does not exist  
         & \multicolumn{2}{|c|}{II: CBV disappears during leptogenesis}
         \\ \hline
         $H_{\mathrm{osc}} < a_m m_{3/2}$ & 
         \multicolumn{2}{|c|}{III: Subtle case} & IV: CBV always exists (CBV problem)
         \\ \hline
    \end{tabular}
    \caption{Four types of the AD leptogenesis scenario. In Case I, the potential does not have a charge-breaking minimum. In Case II, the CBV exists in the potential at zero temperature, but the AD field does not fall into this due to thermal correction. In Case III, although an unwanted global minimum does not exist, the dynamics of the AD field are affected by the unwanted local minimum during leptogenesis.  In Case IV, the AD field falls into the unwanted global minimum. (CBV problem)}
    \label{tab:ADLGscnario}
\end{table}

\section{Results and Implications}
\label{sec:result}

\begin{figure}[tp]
\centering
  {\includegraphics[width=0.9\textwidth]{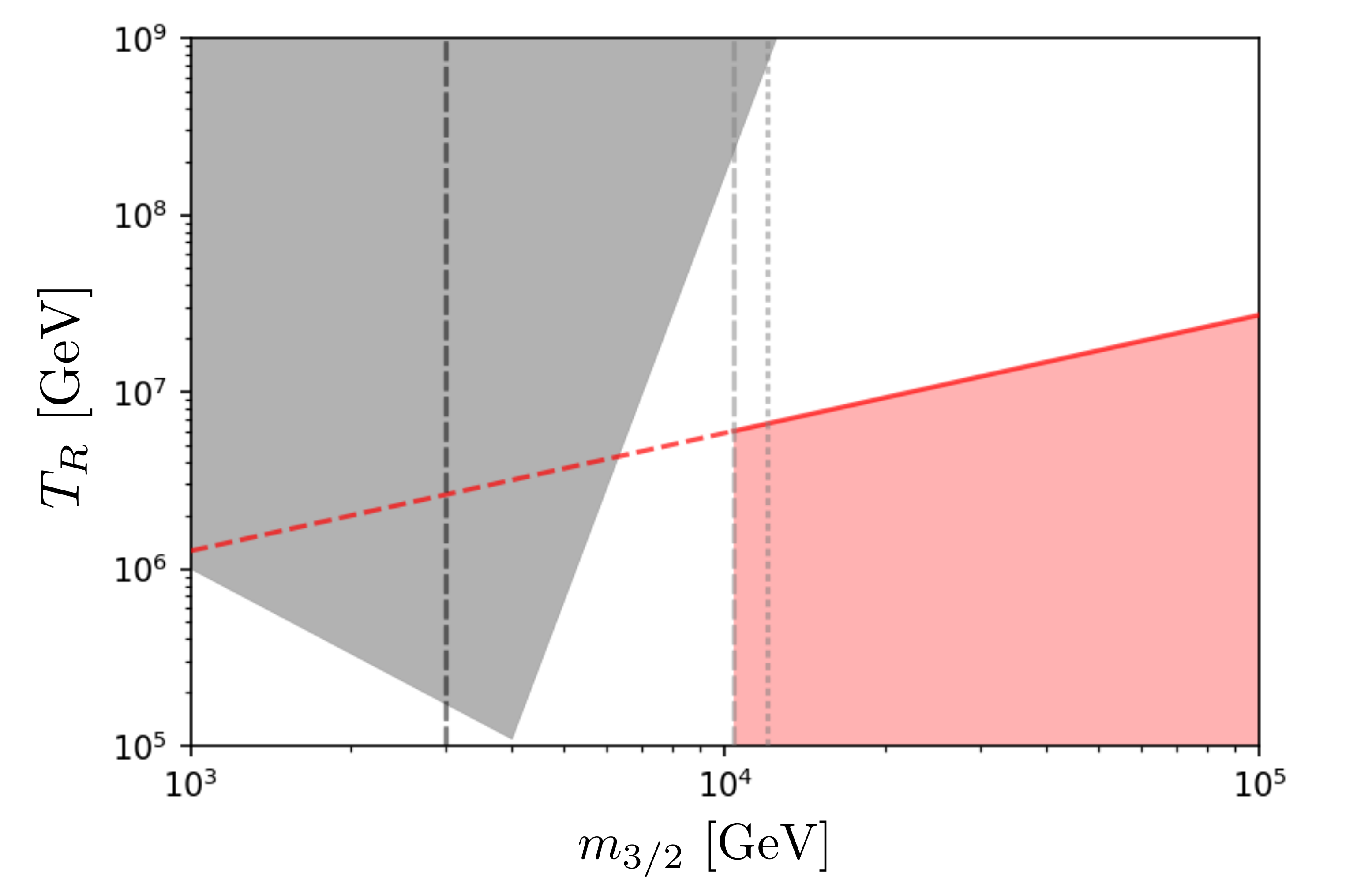}} 
  \caption{
  The allowed regions of the parameter space in the case of AD leptogenesis with the light AD field. We fixed the effective AD field mass at $3~\mathrm{TeV}$ in this figure, which is described by the vertical black dashed line. The vertical gray dashed (dotted) line corresponds to the boundary whether the unwanted local (global) minimum exists or not. The CBV problem can occur when the gravitino mass exceeds the dashed line. The BBN constraint~\cite{Kawasaki:2008qe,Kawasaki:2017bqm} is approximately shown by the gray-shaded region. In the pink-shaded region, the AD leptogenesis does not work due to the trapping at the CBV. Since the boundary of the pink region depends on the effective AD field mass, we show the red dashed line, under which the AD field may fall into the unwanted vacuum depending on the AD field mass.
  }
  \label{fig:Allowedregionin}
\end{figure}

\begin{figure}[tp]
\centering
  {\includegraphics[width=0.9\textwidth]{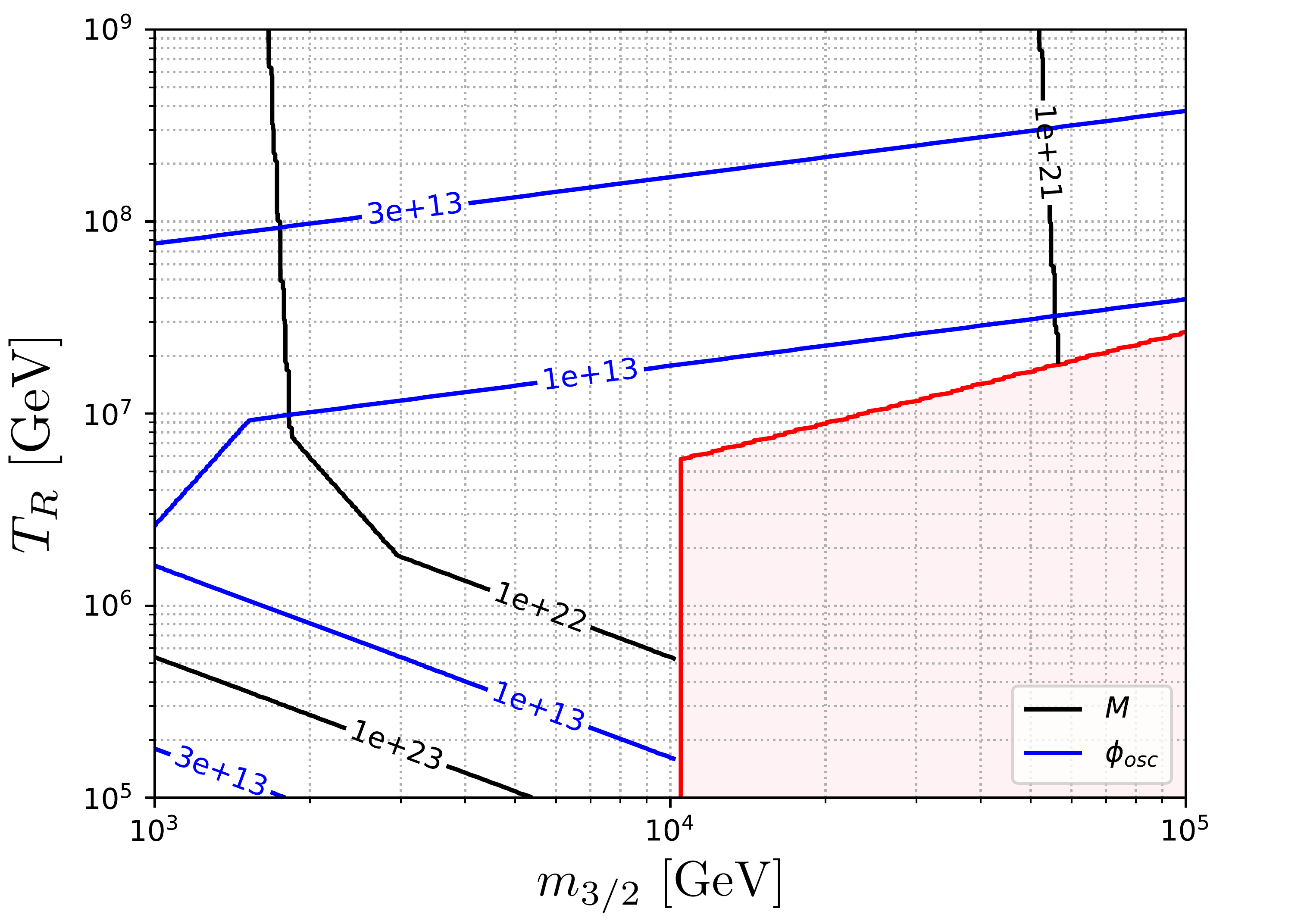}} 
  \caption{The contours of the parameter $M$ that can explain the observed baryon asymmetry in our Universe (blue lines), and the AD field value at the oscillation epoch, $\phi_{\mathrm{osc}}$ (black lines). We took the CP phase $\delta_{\mathrm{CP}}$ to be unity in this figure. The pink region is the same as Fig.~\ref{fig:Allowedregionin} and is forbidden by the CBV problem.
  }
  \label{fig:phioscandcutoff}
\end{figure}

We show our main results in Fig.~\ref{fig:Allowedregionin} and Fig.~\ref{fig:phioscandcutoff}.
In these figures, we assume that the effective AD field mass is $3~\mathrm{TeV}$ during the AD leptogenesis. In Fig.~\ref{fig:Allowedregionin}, it
is described by the vertical black dashed line. 
This is a typical value of the AD field mass, as discussed in Sec.~\ref{subsec:running}. The parameter region 
is divided into two parts: the right side of the gray dashed line, $a_m m_{3/2} > \sqrt{12} ~m_{\phi,\mathrm{eff}} $, and the left side, $a_m m_{3/2} < \sqrt{12} ~m_{\phi,\mathrm{eff}} $, where we have taken $a_m=1$. On the right side, CBV must be hidden by the thermal potential, and therefore the reheating temperature should be sufficiently high. As we will see, the thermal effect mainly comes from the thermal logarithmic contribution, therefore we only focus on this contribution here.
Below the solid red line (in the pink-shaded region), the thermal logarithmic contribution cannot hide the CBV, thus the AD leptogenesis does not work in this region due to the trapping.
In addition, we show the gray dotted line corresponding to $a_m m_{3/2}=4m_{\phi,\mathrm{eff}}$. We do not adopt this line as the boundary of the allowed region, because in the range $\sqrt{12}m_{\phi,\mathrm{eff}} < a_m m_{3/2} < 4 m_{\phi,\mathrm{eff}}$, the AD field would be trapped by unwanted local minimum, which is remaining at zero temperature, and we cannot rely on the usual expression of the produced baryon asymmetry.
Here we have used Eq.~\eqref{TRconstThermalLog}, with $M$ being chosen 
so that the resultant baryon asymmetry is consistent with the present Universe; see Eq.~\eqref{eq:baryonasymmetry} ($a_m$ and the CP phase $\delta_{CP}$ are taken to be unity).
In addition to the constraints to avoid the trapping into the CBV, there are upper bounds on the reheating temperature depending on the gravitino mass, coming from the fact that the gravitino decay may spoil the success of the Big Bang Nucleosynthesis (BBN)~\cite{Khlopov:1984pf,Ellis:1984eq,Kawasaki:2008qe,Kawasaki:2017bqm}. The constraints~\cite{Kawasaki:2008qe,Kawasaki:2017bqm} are approximately shown by the gray-shaded region. 
The AD leptogenesis can successfully work in the remaining white region, 
with an appropriate choice of $M$.

Fig.~\ref{fig:phioscandcutoff} shows the contours of the AD field values at the oscillation epoch, $\phi_{\mathrm{osc}}$ (blue lines), and the parameter $M$ that explains the observed baryon asymmetry of our Universe (black lines). We took the CP phase $\delta_{\mathrm{CP}}$ to be unity in the figure. The pink-shaded region is the same as Fig.~\ref{fig:Allowedregionin}, but we numerically confirmed that this boundary is determined by the thermal logarithmic potential.
The typical value of the AD field is $O(10^{13})~\mathrm{GeV}$ at the oscillation epoch. This implies, together with Eq.~\eqref{deltamphilog}, that  
the choice $m_{\phi,\mathrm{eff}} = 3$ TeV during the AD leptogenesis is consistent with light sleptons at the electroweak scale by the large logarithmic correction.   
The parameter $M$, which is related to the lightest neutrino mass  by Eq.~\eqref{eq:M_and_neutrino_mass}, is typically $O(10^{21-22})~\mathrm{GeV}$. This corresponds to the neutrino mass $m_{\nu1} \sim 3 \times 10^{-(8-9)}~\mathrm{eV}$.\footnote{In our case, the running effect of the neutrino mass is negligible for the following reasons. First, since the RGE for each neutrino mass is proportional to the mass itself~\cite{Babu:1993qv,Chankowski:1993tx},
we expect that the running effect of the lightest neutrino mass is not important at the one loop level. Second, the mass shift from two loop effect is given by~\cite{Davidson:2006tg}
\begin{align}
\label{eq:mass shift}
\Delta m_{\nu1} \sim 10^{-10}~\mathrm{eV}\left(\frac{\tan \beta}{10}\right)^4,
\end{align}
which is smaller than the typical value $m_{\nu1} \sim 3 \times 10^{-(8-9)}~\mathrm{eV}$. If the gravitino is lighter than $O(100)~\mathrm{GeV}$, this shift might not be negligible in particular when $\tan \beta$ is large.}
In the upper region in Fig.~\ref{fig:phioscandcutoff}, the 
parameter $M$ becomes independent of the reheating temperature. This is because for high reheating temperature, $H_{\mathrm{osc}}$ is determined by the thermal logarithm potential, and then the produced baryon asymmetry is independent of the reheating temperature~\cite{Fujii:2001zr}. For a lower reheating temperature, $H_{\mathrm{osc}}$ is determined by the thermal mass or the effective mass $m_{\phi,\mathrm{eff}}$, and the parameter $M$ that can explain the baryon asymmetry depends on the reheating temperature.

Finally, let us discuss the implication of our study to the low-energy SUSY model. If we consider the AD leptogenesis with the low-energy SUSY model, which typically has light sleptons, we need to take care of the CBV problem. In particular, in the model that can explain muon $g-2$ anomaly, light slepton mass leads to light AD field mass, since typically the model has large $\tan\beta$ (cf.~Eq.~\eqref{higgsandmuterm}).
To avoid this problem, the AD field mass should become as large as the gravitino mass due to the running effect, or the reheating temperature should be sufficiently high. Then, the produced baryon asymmetry is determined at a high-energy scale, and therefore almost independent of low-energy SUSY phenomenology at the sub-TeV scale. That is why in our figures, we do not have any parameter dependence related to the low-energy SUSY model.

\section{Summary}
\label{sec:summary}
In this paper, we have studied the 
AD leptogenesis with light slepton. The light slepton, favored by the low-energy SUSY model, can lead to the light AD field mass. Such a light AD field is problematic because it may create a 
CBV.  We have investigated the condition where the CBV exists and found that both thermal correction and quantum correction are important for the (temporal) disappearance of the charge-breaking minimum. Our investigation is generic and independent of the SUSY-breaking mechanism unless
the SUSY-breaking sector affects the scalar potential along the flat direction significantly. We conclude that if the CBV disappears due to the thermal or quantum correction, the correct baryon asymmetry can be produced while avoiding the gravitino problem.


\section*{Acknowledgments}

This work is supported in part by the National Research Foundation of Korea (NRF) grant funded by the Korea government(MSIT) (No. NRF-2021R1A2C2009718 [KE]), the Grant-in-Aid for Innovative Areas (No.19H05810 [KH], No.19H05802 [KH]), Scientific Research B (No.20H01897 [KH]), Scientific Research C (No.19K03842 [KK]) and JSPS KAKENHI Grant (No.22J21260 [JW]).

\bibliographystyle{utphysmod}
\bibliography{ref}

\end{document}